\documentclass[a4paper,11pt]{article}
\pdfoutput=1 

\usepackage{jinstpub} 
\usepackage[separate-uncertainty=true]{siunitx}

\usepackage{comment} 
\usepackage[normalem]{ulem}
\usepackage{graphicx}
\usepackage{float}
\usepackage{subcaption}
\usepackage[version=4]{mhchem}
\usepackage{chemist}

\def\Bcarb{$^{10}$B$_4$C }
\def\Gd{$^{157}$Gd }
\def\He{$^{3}$He }
\def\Li{$^{6}$Li }
\def\LiSete{$^{7}$Li }
\def\Cd{$^{113}$Cd }
\def\B{$^{10}$B}

\DeclareMathOperator\erfc{erfc}

\title{Double-GEM based thermal neutron detector prototype}

\author[a,1]{L. A. Serra Filho\note{Corresponding author.},}
\author[a]{R. Felix dos Santos,}
\author[a]{G. G. A. de Souza,}
\author[a]{M. M. M. Paulino,}
\author[b]{F. A. Souza,}
\author[b]{M. Moralles,}
\author[c]{H. Natal da Luz,}
\author[a]{M. Bregant,}
\author[a]{M. G. Munhoz,}
\author[d,e]{Chung-Chuan Lai,}
\author[d,f]{Carina Höglund,}
\author[d]{Per-Olof Svensson,}
\author[d]{Linda Robinson}
\author[d,g]{and Richard Hall-Wilton}

\affiliation[a]{Instituto de Física da Universidade de São Paulo\\
Rua do Matão 1371, 05508-090 Cidade Universitária, São Paulo, Brasil}
\affiliation[b]{Instituto de Pesquisas Energéticas e Nucleares,\\Avenida Lineu Prestes 2242, 05508-000 Cidade Universitária, São Paulo, Brasil}
\affiliation[c]{Institute of Experimental and Applied Physics, Czech Technical University in Prague\\ Husova 5, 110 00 Prague 1, Czech Republic}
\affiliation[d]{Detector Group, European Spallation Source ERIC (ESS)\\
European Spallation Source ERIC (ESS), P.O. Box 176, 221 00 Lund, Sweden}
\affiliation[e]{Linköping University\\
Linköping University, 581 83 Linköping, Sweden}
\affiliation[f]{Current Affiliation: Impact Coatings AB\\
Westmansgatan 29G, 582 16 Linköping, Sweden}
\affiliation[g]{Universit\`a degli Studi di Milano-Bicocca\\
Piazza della Scienza 3, 20126 Milano, Italy}

\emailAdd{lserra@if.usp.br}

\abstract{The Helium-3 shortage and the growing interest in neutron science constitute a driving factor in developing new neutron detection technologies. In this work, we report the development of a double-GEM detector prototype that uses a \Bcarb layer as a neutron converter material. GEANT4 simulations were performed predicting an efficiency of \SI{3.14(10)}{\percent}, agreeing within $2.7\,\sigma$ with the experimental and analytic detection efficiencies obtained by the detector when tested in a \SI{41.8}{\milli \electronvolt} thermal neutron beam. 
The detector is position sensitive, equipped with a \num{256}+\num{256} strip readout connected to resistive chains, and achieves a spatial resolution better than \SI{3}{\milli\meter}. The gain stability over time was also measured with a fluctuation of about \SI{0.2}{\percent\per\hour} of the signal amplitude. A simple data acquisition with only \num{5} electronic channels is sufficient to operate this detector.}

\keywords{Gaseous detectors; Neutron detectors (cold, thermal, fast neutrons); Micropattern gaseous detectors (MSGC, GEM, THGEM, RETHGEM, MHSP, MICROPIC, MICROMEGAS, InGrid, etc)}

\arxivnumber{2205.07122} 

\begin{document}
\maketitle
\flushbottom

\section{Introduction}
Several advances in scattering techniques inspired the interest in neutron science, which provides different applications in areas such as chemistry, physics, biology, medicine and engineering research. These advances are strongly dependent on detector developments, which has also grown in the last years, along with the capability to obtain bright monochromatic neutron sources, whether in large or small facilities \cite{PIETROPAOLO20201, HALLWILTON2014, Ledoux2021, DIGIOVINE2021}.

The \He proportional counter is one of the most widely used type of neutron detectors, still widely used. Beyond its historical tradition, it presents several advantages, such as the large neutron capture cross-section ($\sigma =$ \SI{5325}{\barn} for \SI{25}{\milli\electronvolt} \cite{tendl}), low sensitivity to gamma-rays, and the possibility of working on a wide neutron energy range in the low energy region \cite{PIETROPAOLO20201}. However, due to a shortage of this gas since the early 2000s \cite{He3shortage}, the scientific community keeps searching for alternative solutions with $\ce{^{3}_{}He}$-free detectors \cite{Sacchetti2015}, employing \mbox{\Cd,} \Gd, \B\mbox{}, and \Li \cite{MIYAKE2011, PIETROPAOLO2013, Schooneveld_2016, COMPAAN2020, ALBANI2020} as neutron converters. Nevertheless, substituting \He has the price of reducing the neutron detection efficiency or increasing the detection complexity.

One versatile alternative for neutron detection consists of using one of the referred neutron converter isotopes with the Gas Electron Multiplier (GEM) \cite{Sauli1997}, a microstructure widely used to detect charged particles that can cover large sensitive areas,  present fair energy and position resolutions as well as robustness. These aspects consolidated these microstructures, which are already in use in several applications, such as high energy physics \cite{ALICE, COMPASS, ABBANEO2017298}, muon tomography \cite{Muon_tomography}, X-rays fluorescence imaging \cite{geovoXRF}, and, more recently, neutron detection \cite{KLEIN2011, KOHLI2016, ZHOU2020, ZHOU2021}.

This work presents a double-GEM detector which uses \Bcarb as a neutron converter, deposited on its aluminum cathode. The ionizing products generated by the \B(n,$\alpha$)\LiSete capture reaction, shown in Eq.~\ref{eq:decays}, are a critical factor in detector's design. Given that \mbox{\B } has a relatively large neutron capture cross section of \SI{3870}{b} for \SI{25}{\milli\electronvolt} \cite{tendl} and about \SI{19}{\percent} natural abundance \cite{elements-abundance}, one of its compounds, $^{10}$B$_4$C, is particularly interesting for our purposes, considering it also presents good wear resistance besides chemical and thermal stability \cite{ESSref3}.

\begin{ChemEquation}
  n + \ce{^{10}}B
  \reactrarrow{1pt}{1cm}{}{}
  \begin{cases}
  \ce{^{7}Li}\,(\SI{840}{\kilo\electronvolt}) + \ce{^{4}}He\,(\SI{1470}{\kilo\electronvolt}) + \gamma\, (\SI{480}{\kilo\electronvolt}) \mbox{   }(94\%)\\
  \ce{^{7}}Li\,(\SI{1015}{\kilo\electronvolt})  + \ce{^{4}}He\,(\SI{1775}{\kilo\electronvolt})    \mbox{   }(6\%)
  \end{cases}
  \label{eq:decays}
\end{ChemEquation}

This prototype can be easily assembled and offers the possibility to cover large areas, an essential requirement for some applications. The detector was tested in the IEA-R1 nuclear research reactor \cite{PERROTTA1998}, at the Nuclear and Energy Research Institute (IPEN) in S\~ao Paulo, Brazil, using a monochromatic thermal neutron beam. Its versatility resulted from quickly obtaining neutron position sensitivity using a few electronics channels and presenting gain stability. 

\section{Experimental setup}
\label{sec:experimental}
Our double-GEM detector is composed of a stack with a \SI{0.5}{\milli\meter} thick aluminum cathode coated with amorphous enriched boron carbide (\Bcarb), followed by two GEM foils, provided by CERN.

The deposition of the \Bcarb  film on the Al cathode was done by direct-current magnetron sputtering in ESS Linköping Detector Coatings Workshop, located at Linköping, Sweden. The facility is equipped with an industrial deposition unit (CC800/9, CemeCon AG) and has been dedicated to the development and production of B$_4$C coatings as the solid converting layers in neutron detectors. Details about the deposition machine and characterization of the B$_4$C thin films for neutron detectors can be found in \cite{ESSref1,ESSref2,ESSref3}.

\begin{figure}[H]
    \centering
    \includegraphics[width=0.4\textwidth]{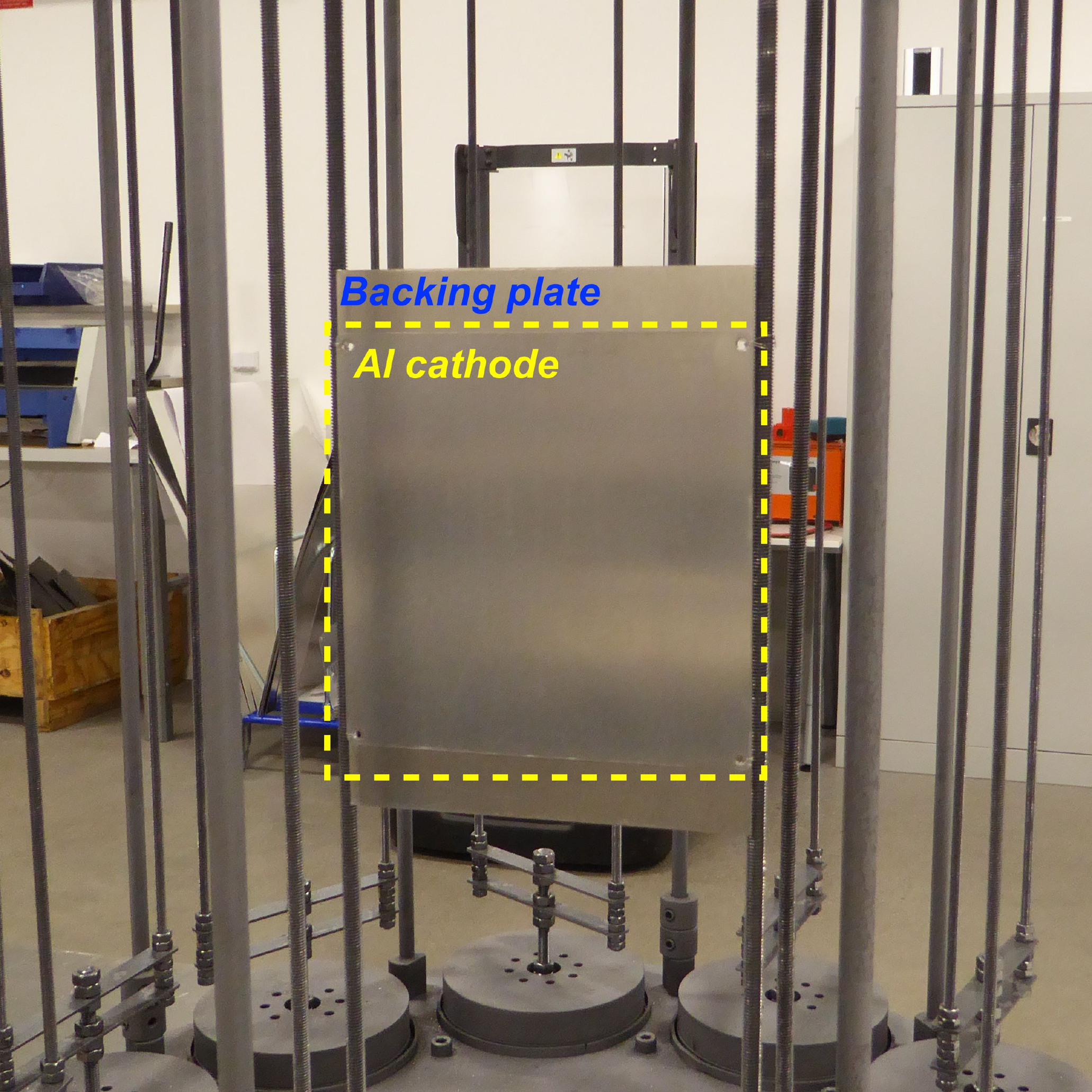}
    \caption{The assembly of the Al cathode plate (highlighted with the dotted box) on the Al backing plate for deposition.}
    \label{fig:ESS}
\end{figure}

In this work, two \Bcarb targets (\B\mbox{ }enrichment $>$ 95 wt.\%) were installed on sputtering cathodes in the deposition machine. The Al plate was attached to another Al backing plate to achieve \Bcarb coating on only one surface, where the whole assembly was mounted to an electrically grounded sample table (see Fig. \ref{fig:ESS}). The system was first pumped down to a primary vacuum pressure of  \SI{1.5e-4}{Pa} (\SI{1.1e-6}{Torr}) while the Al plate was heated up to  \SI{280}{\degreeCelsius} for \SI{3}{\hour} mainly to remove the water residuals on the surfaces. After an additional \SI{2}{\hour} cooling, the plate was then treated with radio frequency plasma etching in \SI{0.35}{Pa} of pure Ar to improve the coating’s adhesion by removing native oxides on the substrate surface and increasing surface roughness. Thereafter, the system was filled up with \SI{0.8}{Pa} (\SI{6}{mTorr}) of Ar before the beginning of the \Bcarb deposition. During the deposition, the sample table was continuously rotated to obtain a better thickness uniformity. A polished Si reference was mounted at an equivalent position of the Al plate in the same deposition run for thickness measurement of the coating and the thickness of \Bcarb coating was measured to be approximately \SI{2.2}{\micro\metre} by a profilometer (DektakXT, Bruker).

\begin{figure}[h]
    \centering
    \includegraphics[width=0.6\textwidth]{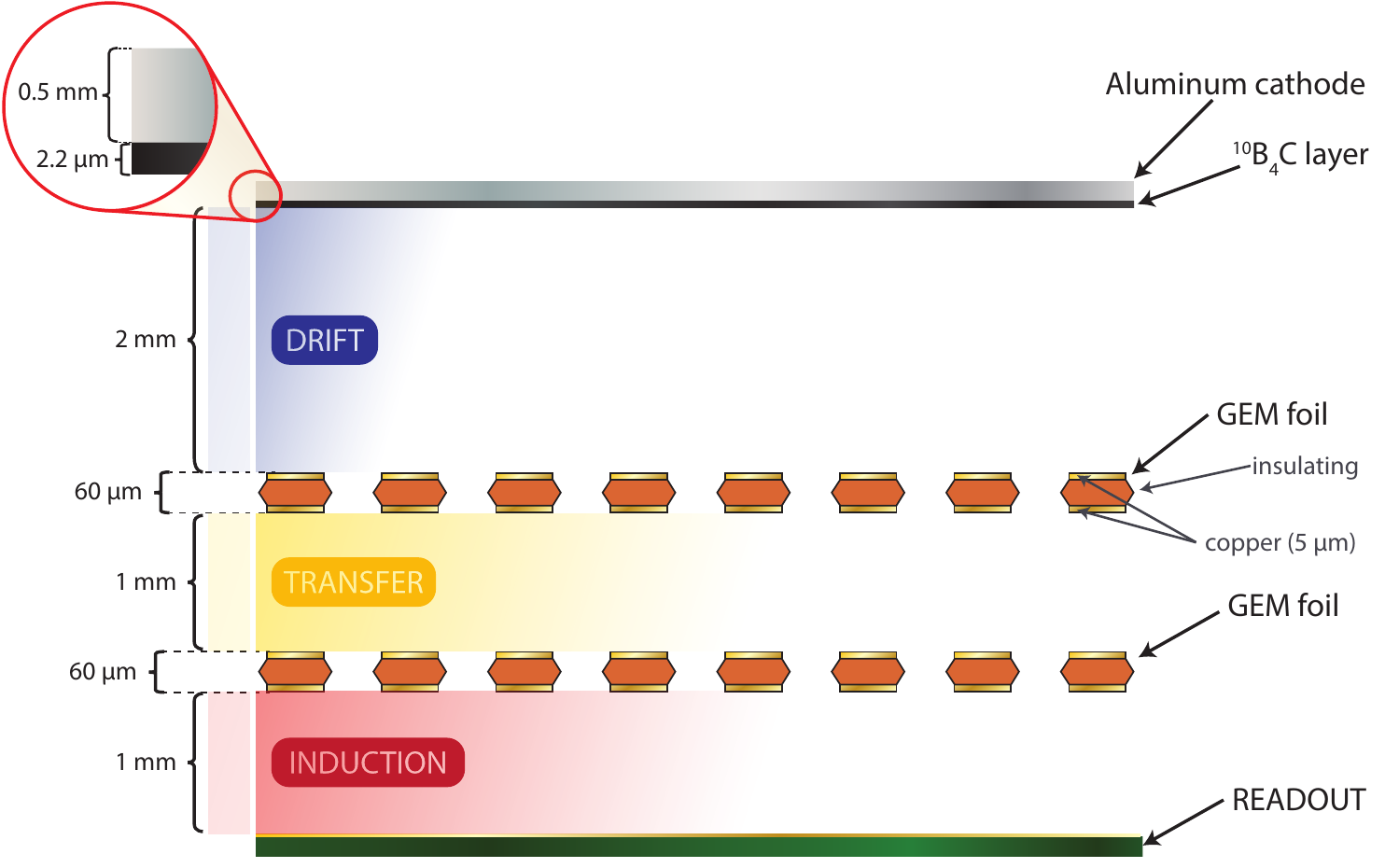}
    \caption{Detector scheme, showing the drift, transfer and induction regions. Not to scale. Figure adapted from \cite{LucasMaster}.}
    \label{fig:scheme}
\end{figure}

In addition to the deposition of the converter layer, the geometrical disposition of the elements within the detector is a key factor regarding the signal we want to measure. The drift, transfer, and induction regions of our prototype are represented in Fig. \ref{fig:scheme}. The drift region has \SI{2}{\milli\meter} thickness and is biased with \SI{100}{V}, while the transfer and induction regions have both \SI{1}{\milli\meter} thickness and are biased with \SI{300}{V} and \SI{400}{V}, respectively. The detector works under an open flow of an Ar/CO$_2$ (90/10) gas mixture, with a flow of \SI{6}{L/h} through the \SI{0.85}{\liter} volume aluminum chamber. The readout plane, produced at CERN, consists of \num{256}+\num{256} strips in X and Y directions, covering a \SI{100}{\centi\meter\squared} sensitive area.

The thickness of the drift region determines the detector's sensitivity to background gamma rays. In this context, a common requirement for thermal neutron detectors is to be insensitive to these gammas.
Preliminary tests comparing \SI{20}{\milli\metre} and \SI{2}{\milli\metre} drift regions have shown that the latter detected five times fewer background events (approximately \SI{3}{\hertz}), measured with the neutron beam shutter closed. The share of events within the background range varied from \SI{56}{\percent} for the \SI{20}{\milli\metre} drift region to \SI{18}{\percent} for the \SI{2}{\milli\metre} region. These tests were taken into account to determine the current \SI{100}{\kilo\electronvolt} threshold, which allows the detector to operate above the gamma contamination region. Therefore, we operated the detector using the \SI{2}{\milli\metre} drift region, which is possible because the reaction products' high ionization power generates a large number of primary electrons. That way, even a tiny segment of their tracks is enough to provide a significant signal.

\begin{figure}[]
    \centering
    \includegraphics[height=0.3\textwidth]{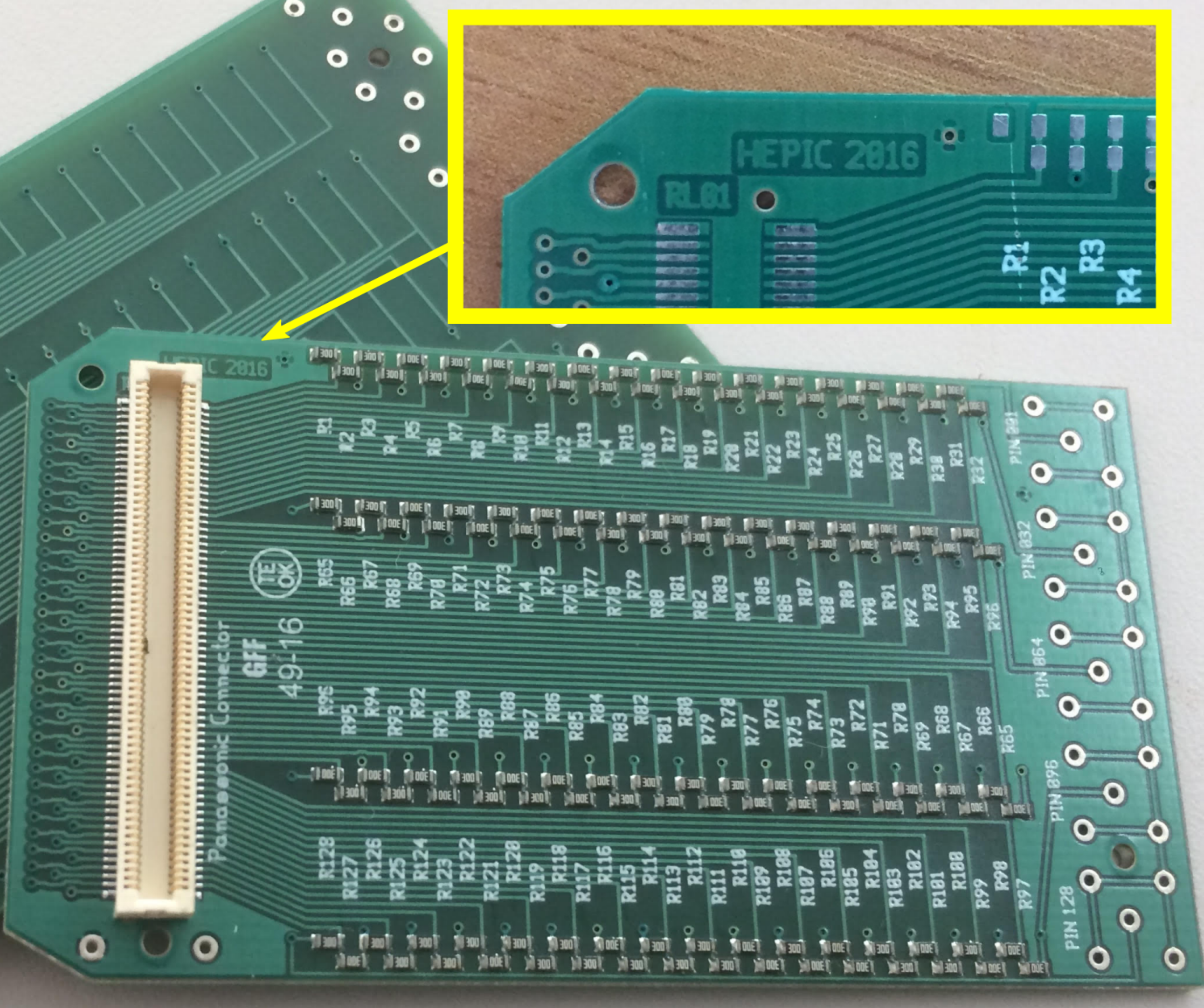}
    \caption{Printed Circuit Board (PCB) resistive chain to fit the detector signal output. Each board has 128 SMD resistors and a PANASONIC\textregistered 130-pin connector. Since the readout has 256 strips for each direction, two plates are used for each direction. Figure from \cite{geovo}.}
    \label{fig:rchain_scheme}
\end{figure}

\begin{figure}[]
    \centering
    \includegraphics[width=0.9\textwidth]{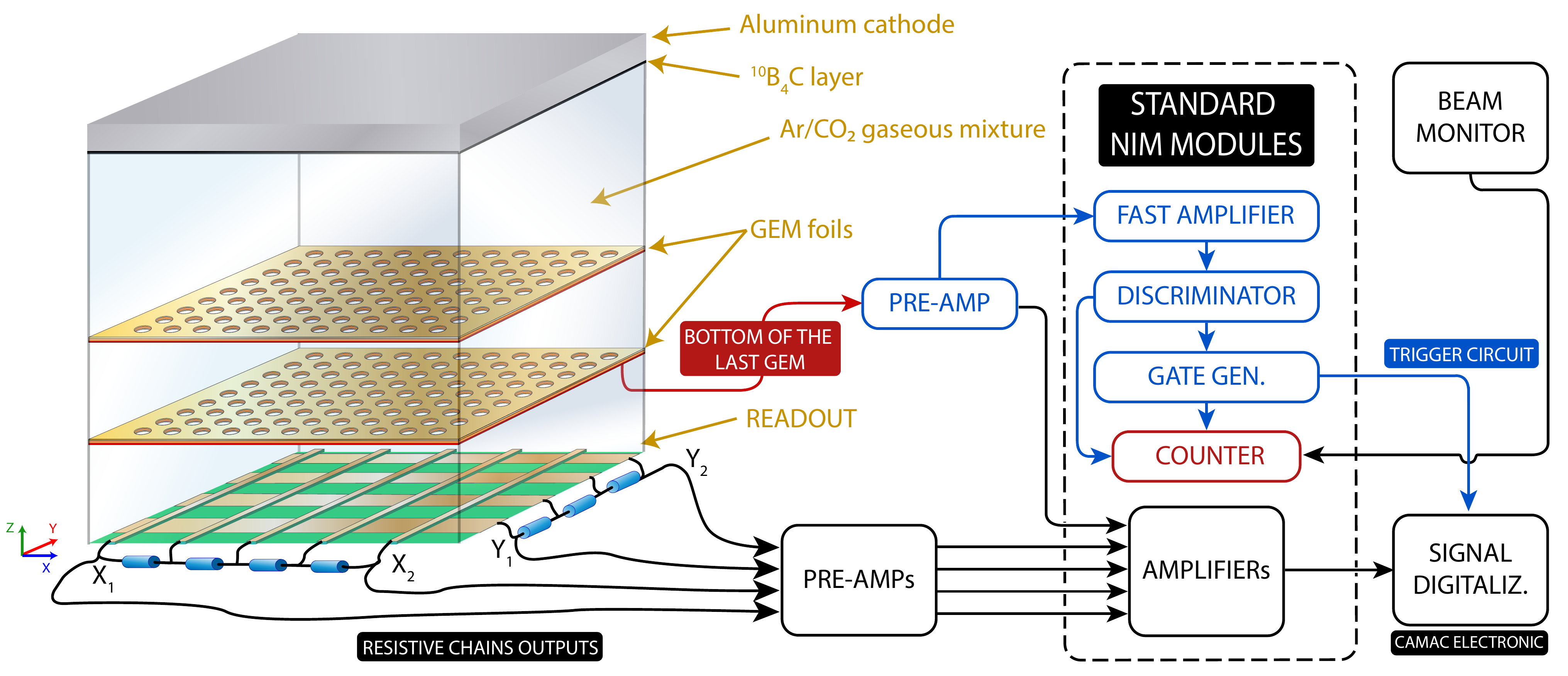}
    \caption{Signal acquisition setup scheme. The trigger circuit is pictured in blue. Figure adapted from \cite{LucasMaster}.}
    \label{fig:eletronic_scheme}
\end{figure}

The signal multiplied by the GEM stack is collected by \num{256}+\num{256} strips, which are interconnected by a resistive chain composed of \SI{60}{\ohm{}} Surface Mount Device (SMD) resistors with \SI{0.1}{\percent} precision, as shown in Fig. \ref{fig:rchain_scheme}. There are five signals collected to reconstruct a neutron image: two signals for the X direction, two for the Y direction and the charge signal, collected from the bottom electrode of the last GEM. The charge signal is used as the trigger for the data acquisition, as shown in Fig. \ref{fig:eletronic_scheme}, since it is position-independent, coming directly from the multiplication. Considering the expected extraction efficiency $\epsilon_{\text{ext}}$ for the last GEM \cite{geovo}, the charge signal collected at its bottom is given by $1-\epsilon_{\text{ext}}$, which is approximately \SI{64}{\percent} of the multiplied charge for our configuration.

In order to reconstruct the event's position, we use Eq.~\ref{eq:charge_dist} \cite{geovoXRF}, where $Q$ is the charge information collected from the bottom of the last GEM, and $L_x$ and $L_y$ are the lengths of the sensitive area for each direction. As indicated in Fig. \ref{fig:eletronic_scheme}, $X_1$ and $X_2$ are the signals collected at each end of the resistive chain in the X direction, while $Y_1$ and $Y_2$ are the equivalent signals in the Y direction.
\begin{equation}
\label{eq:charge_dist}
    X_{*} = L_x\frac{X_2 - X_1}{Q} \qquad Y_{*} = L_y\frac{Y_2 - Y_1}{Q}
\end{equation}

\begin{figure}[]
    \centering
    \includegraphics[width=0.7\textwidth]{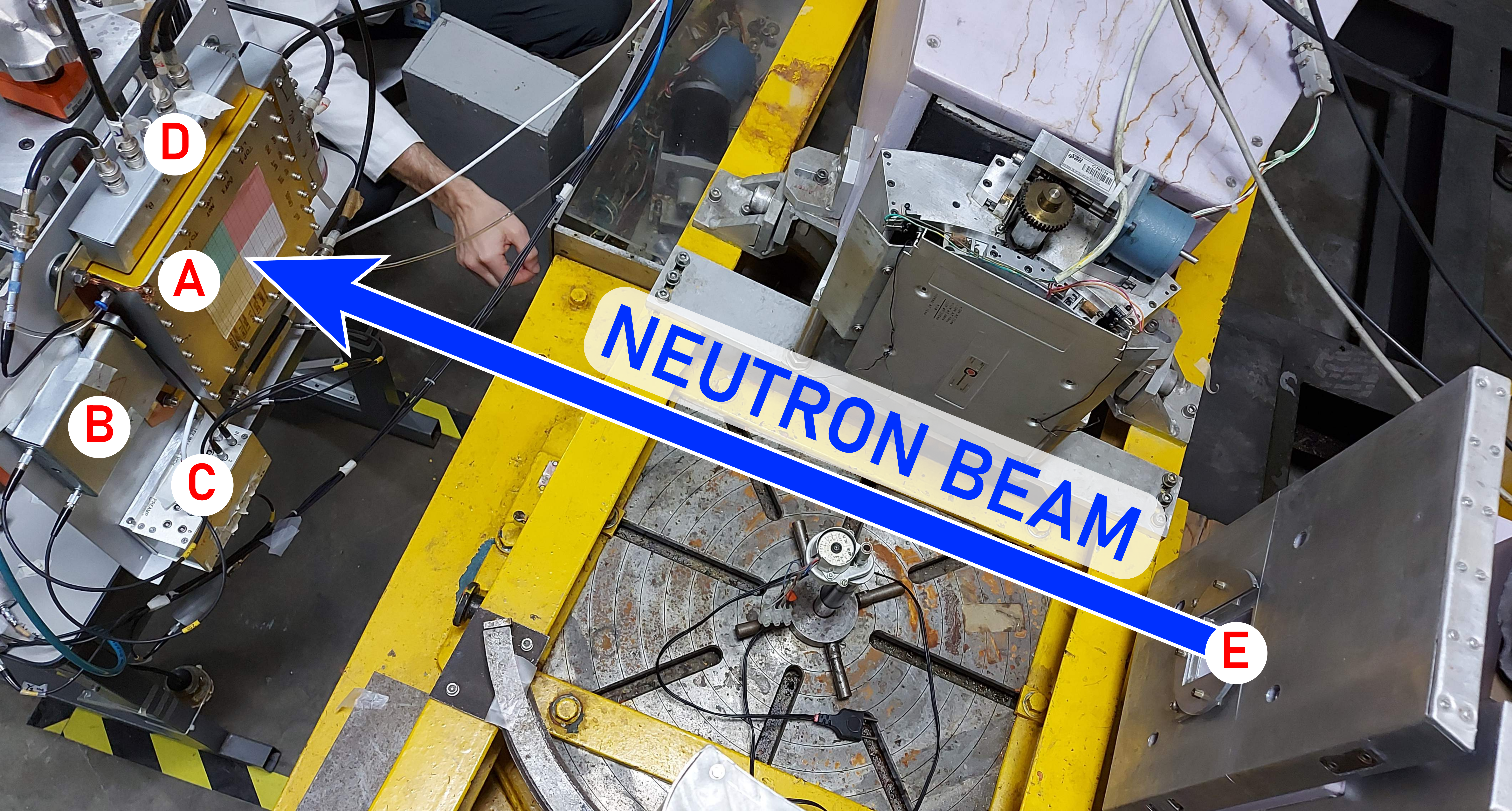}
    \caption{Position of the detector (A) relatively to the beam. The following elements are also indicated: ground shielding of the y-direction resistive chains (B), preamplifiers box (C), high voltage inputs (D), and the beam outlet (E).}
    \label{fig:beam_detector}
\end{figure}

As mentioned before, the detector was tested at the IEA-R1 nuclear reactor using the same neutron beam of the AURORA diffractometer \cite{PARENTE2010678} (see Fig. \ref{fig:beam_detector}). This monochromatic neutron beam is selected out of a collimated radial neutron beam from the core of the reactor, which operates at \SI{4.5}{\mega\watt}. A multi-blade focusing silicon monochromator is used to select these neutrons with a fixed wavelength of \SI{1.399}{\angstrom} (\SI{41.80}{\milli \electronvolt}) providing a \SI{6.22(19)e4}{n\, \centi\metre^{-2}\second^{-1}} flux ($\phi$), measured by the Nuclear Metrology Laboratory at IPEN with the $^{197}\mathrm{Au}(n,\gamma)^{198}\mathrm{Au}$ reaction, using a \SI{12}{\milli\metre} diameter thin gold foil, centered at the beam outlet (letter E of Fig. \ref{fig:beam_detector}).

The detector (letter A of Fig. \ref{fig:beam_detector}) and the preamplifiers' box (letter C of Fig. \ref{fig:beam_detector}) were mounted over an aluminum plate. This setup was fixed to a X-Y-Z precision table, that allowed the fine tuning of the detector's position with respect to the beam.

\section{Results and discussion}
In this section, we present the main experimental results concerning the characterization of the detector: the spatial resolution obtained by two different methods, the detector efficiency, and the gain stability. We also discuss how these results relate to the previous expectations obtained with simulations.

\subsection{Position calibration and resolution}
\begin{figure}[H]
\begin{subfigure}[T]{.5\textwidth}
  \centering
  \includegraphics[width=0.65\linewidth]{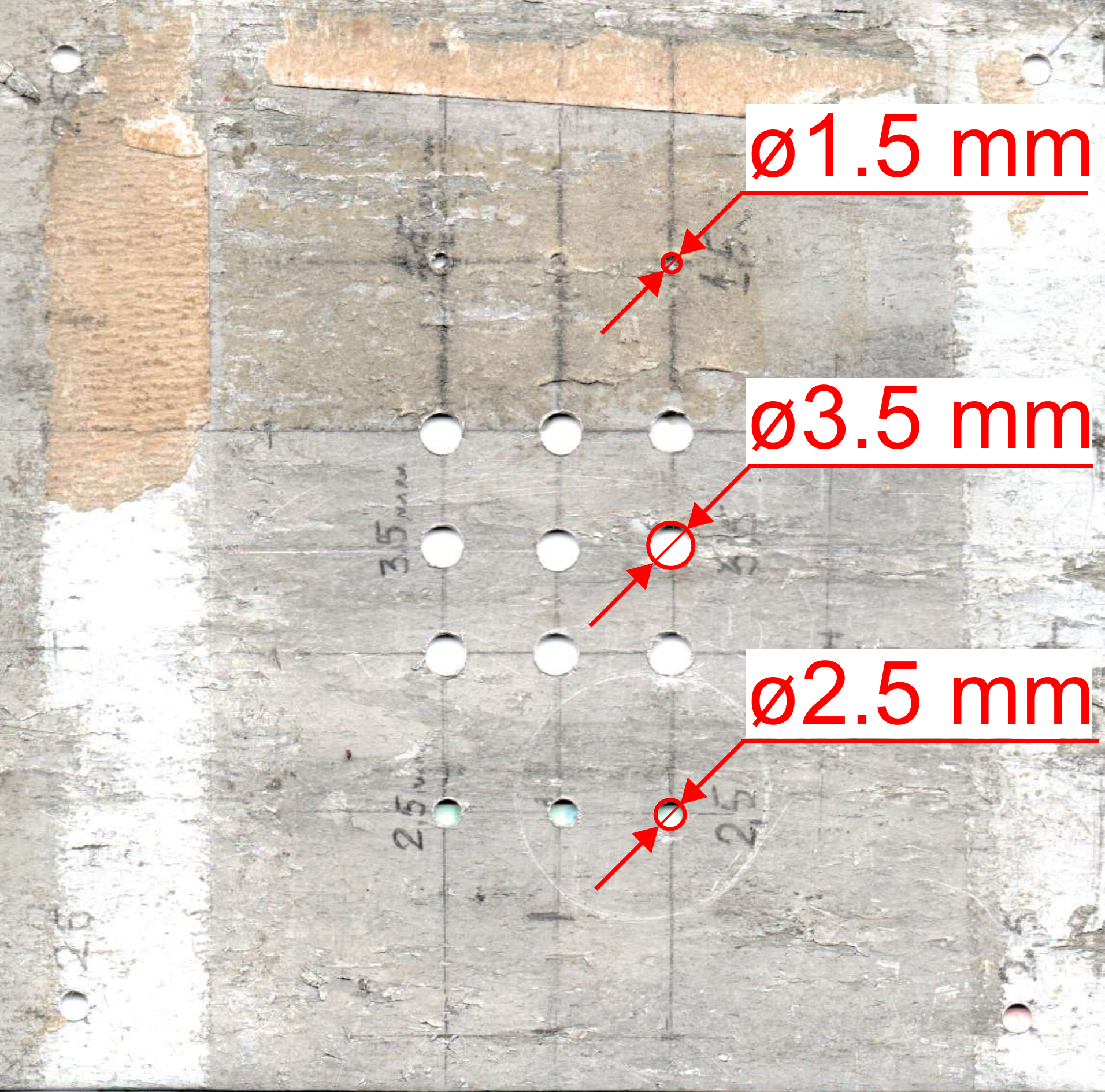}
  \label{fig:calib1}
\end{subfigure}%
\begin{subfigure}[T]{.5\textwidth}
  \centering
  \includegraphics[width=0.8\linewidth]{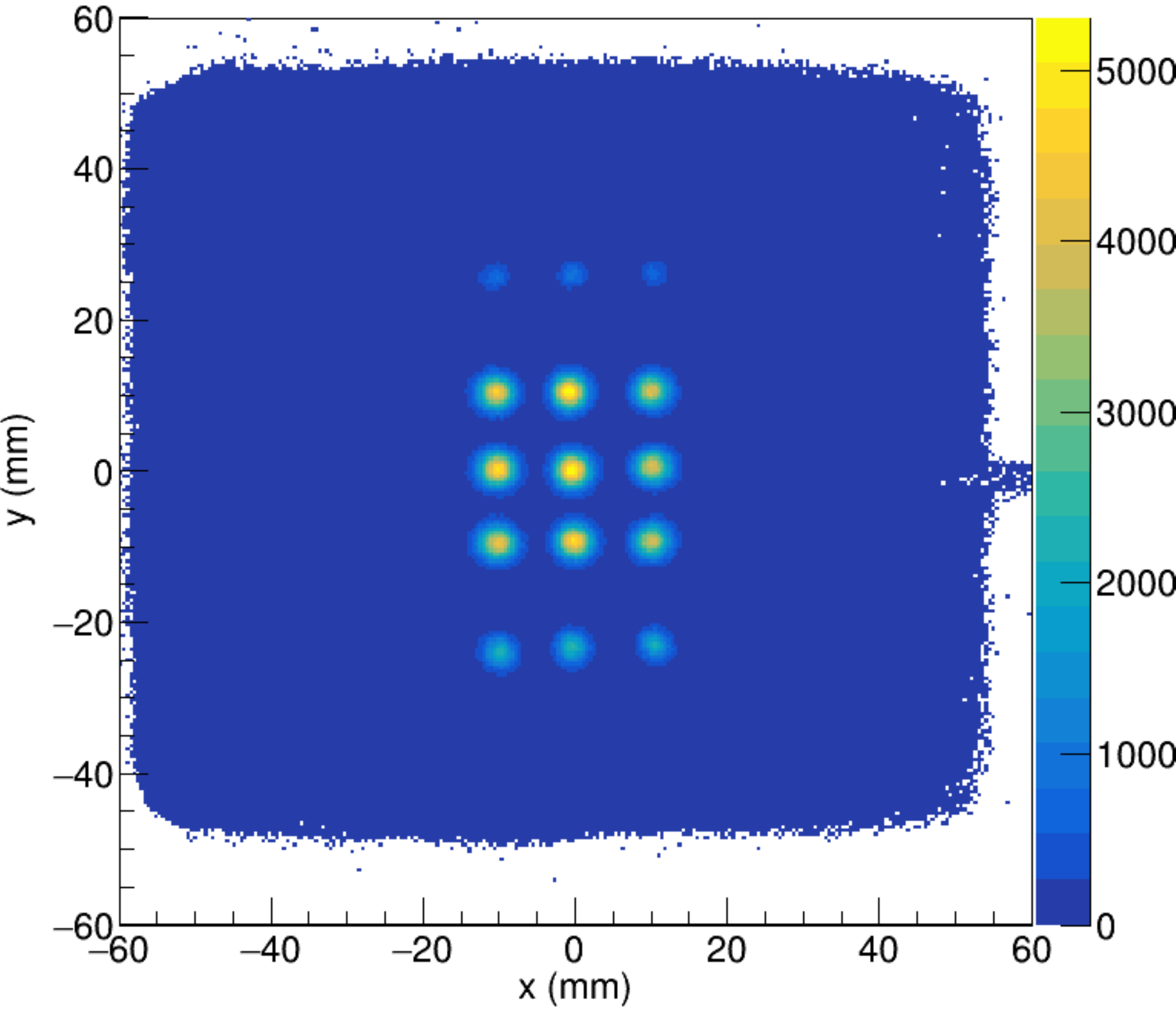}
  \label{fig:calib2}
\end{subfigure}
\caption{Left: digital image of the mask obtained with a table scanner. Right: calibrated neutron image.}
\label{fig:calibration}
\end{figure}

To perform the spatial calibration of the detector, we used handmade cadmium masks which has a high neutron absorption cross section. Holes of different diameters were drilled in three masks. The positions and areas of the holes were measured using an image obtained with a table scanner. One of the produced masks in shown in Fig. \ref{fig:calibration}, with its \num{15} holes, and its neutron image, obtained with the mask centered with respect to the detector's sensitive area. 

\begin{figure}[]
\centering
    \begin{subfigure}{.4\textwidth}
      \centering
      \includegraphics[width=1\linewidth]{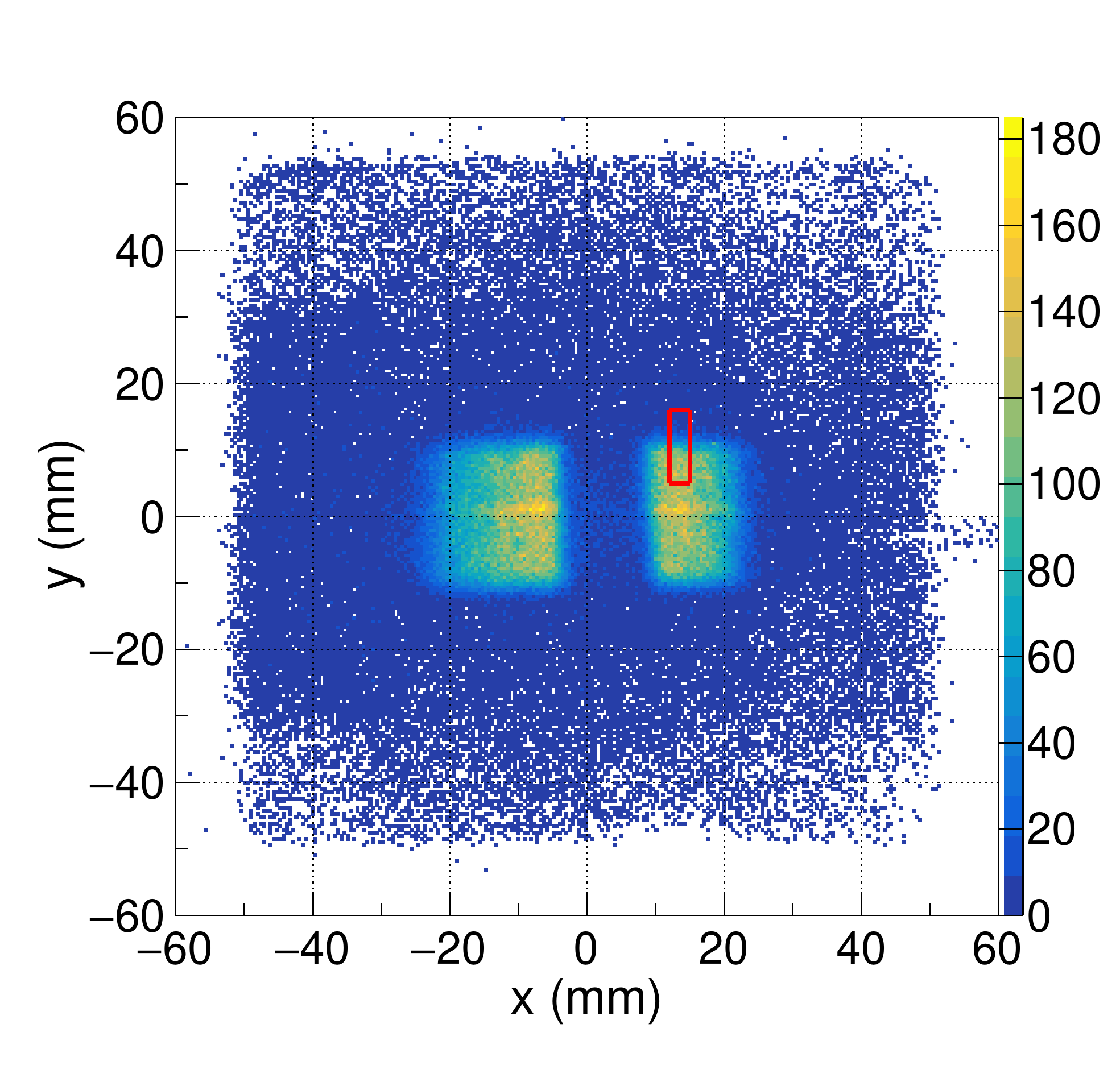}
    \end{subfigure}
    \begin{subfigure}{.4\textwidth}
      \centering
      \includegraphics[width=1\linewidth]{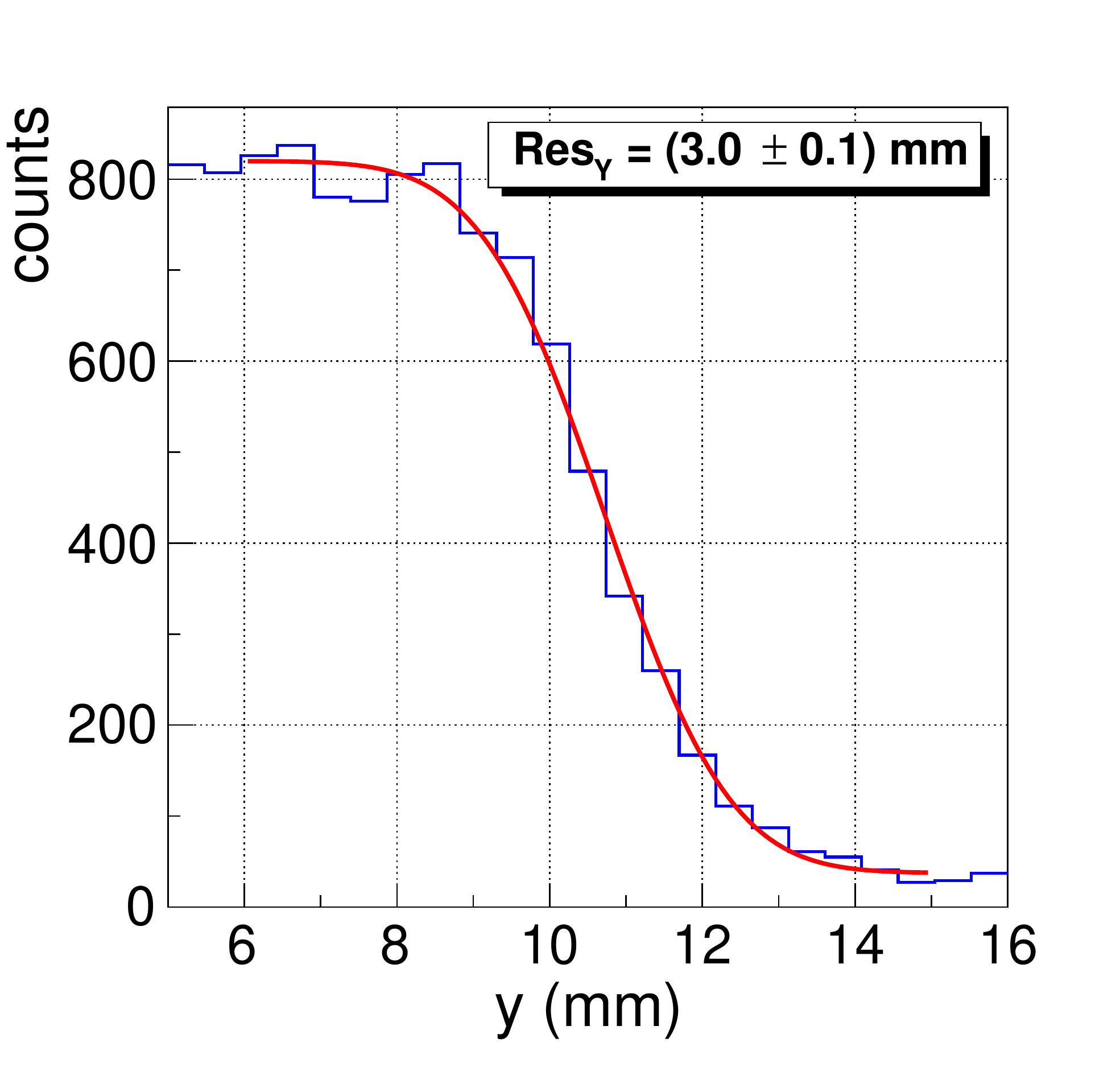}
    \end{subfigure}
\caption{Left: neutron image of one of the masks, with a selected region of an edge marked as the red rectangle. Right: Complementary error function ($\erfc$) adjusted in the y-projection of the chosen area, the spatial resolution obtained is \SI{3.0(1)}{\milli\metre}. Figure adapted from \cite{LucasMaster}.}
\label{fig:edges}
\end{figure}

The calibration is given by a linear dependency between the coordinates of certain holes in arbitrary units and their coordinates in known distances. The chosen holes are colinear in the X and Y directions, normally the ones in the central column for Y direction and the central row for X direction. We measured the detector spatial resolution by two methods: edge and point spread functions. The first one consists of using edges to fit step functions, as shown in Fig. \ref{fig:edges}.

The data was fitted with the complementary error function
\begin{equation}
    \erfc(p) = 1 - \frac{2}{\sqrt{\pi}} \int_{0}^{p} e^{-t^{2}}dt,
    \label{eq:erf_function}
\end{equation}
where
\begin{equation}
    p(x) = \frac{x-\mu}{\sqrt{2}\sigma}.  
    \label{eq:PvarErf_function}
\end{equation}

This function provides the same parameters of interest ($\sigma$ and $\mu$) as the edge spread function (ESF). In this work, we define the position resolution as the FWHM of the Gaussian (or the 10-to-90\,\% distance of the ESF), i. e., 2.35$\sigma$.

\begin{figure}[]
    \centering
    \includegraphics[width=0.5\textwidth]{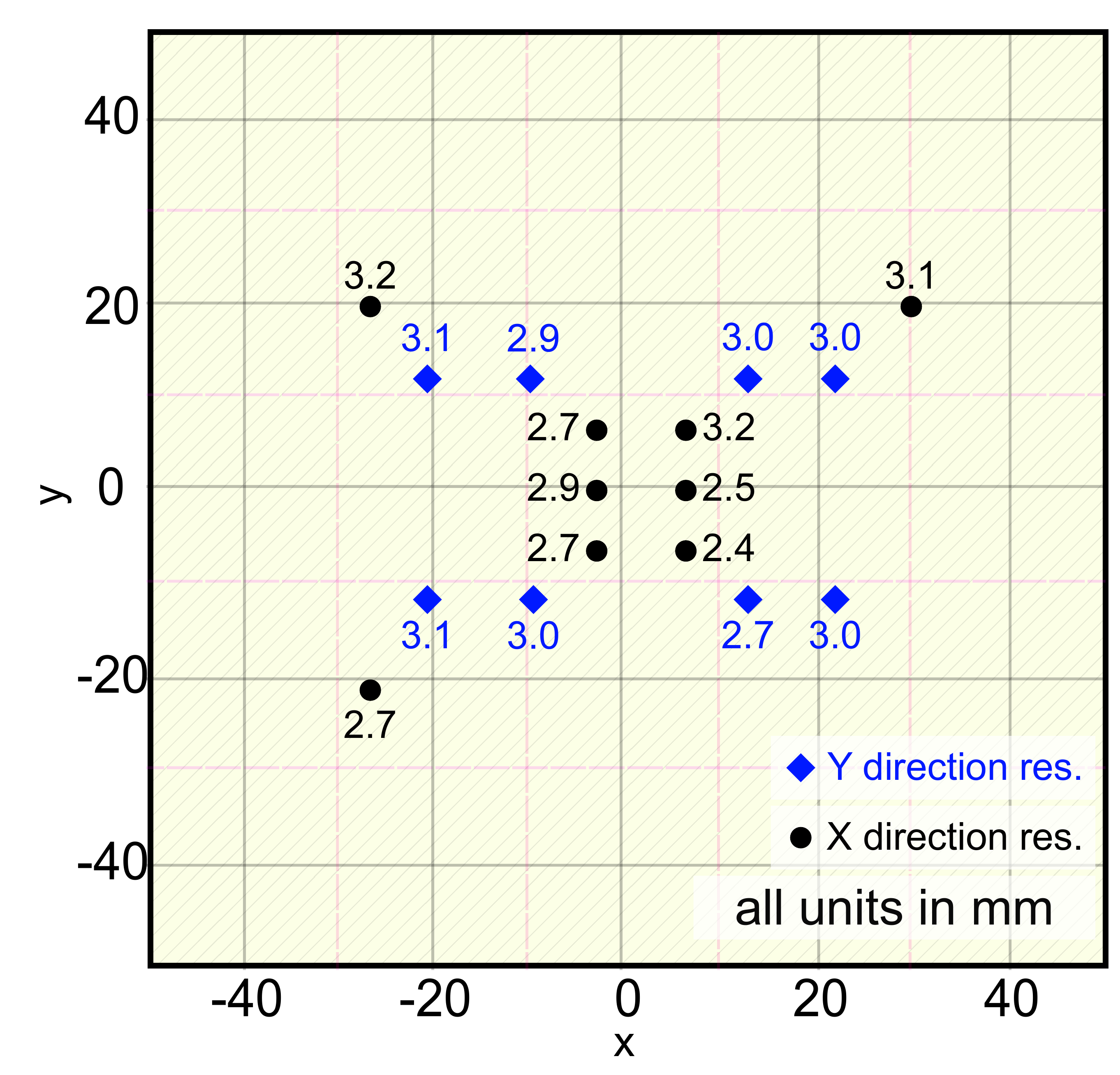}
    \caption{The points where the spatial resolution was measured using the edge spread function. The set of spatial resolution values resulted in an average value of \SI{2.8(3)}{\milli\metre} for the X direction and \SI{3.0(1)}{\milli\metre} for the Y direction, compatible with each other. We considered the uncertainties of the average spatial resolution for each direction as the standard deviation of each set (X and Y). Consequently, it is reasonable to consider the detector's resolution as the average of the resolutions for both directions, which results \SI{2.9(2)}{\milli\metre}.}
    \label{fig:edgesmap}
\end{figure}

We measured the position resolution with the edge spread function in different regions of the detector's sensitive area and along X and Y coordinates. The results are shown in Fig. \ref{fig:edgesmap}.

\begin{figure}[]
\begin{subfigure}{.5\textwidth}
  \centering
  \includegraphics[width=1\linewidth]{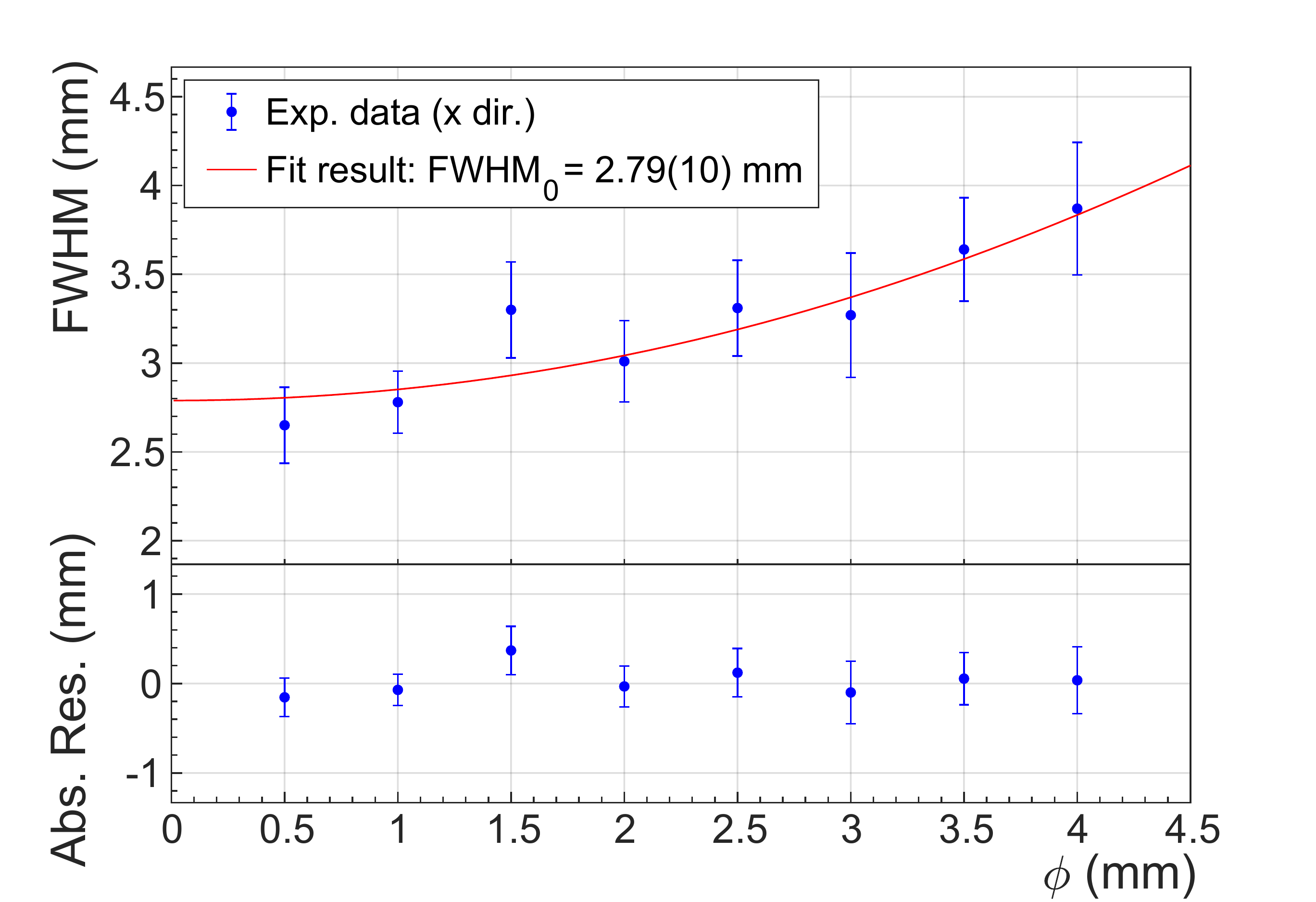}
\end{subfigure}
\begin{subfigure}{.5\textwidth}
  \centering
  \includegraphics[width=1\linewidth]{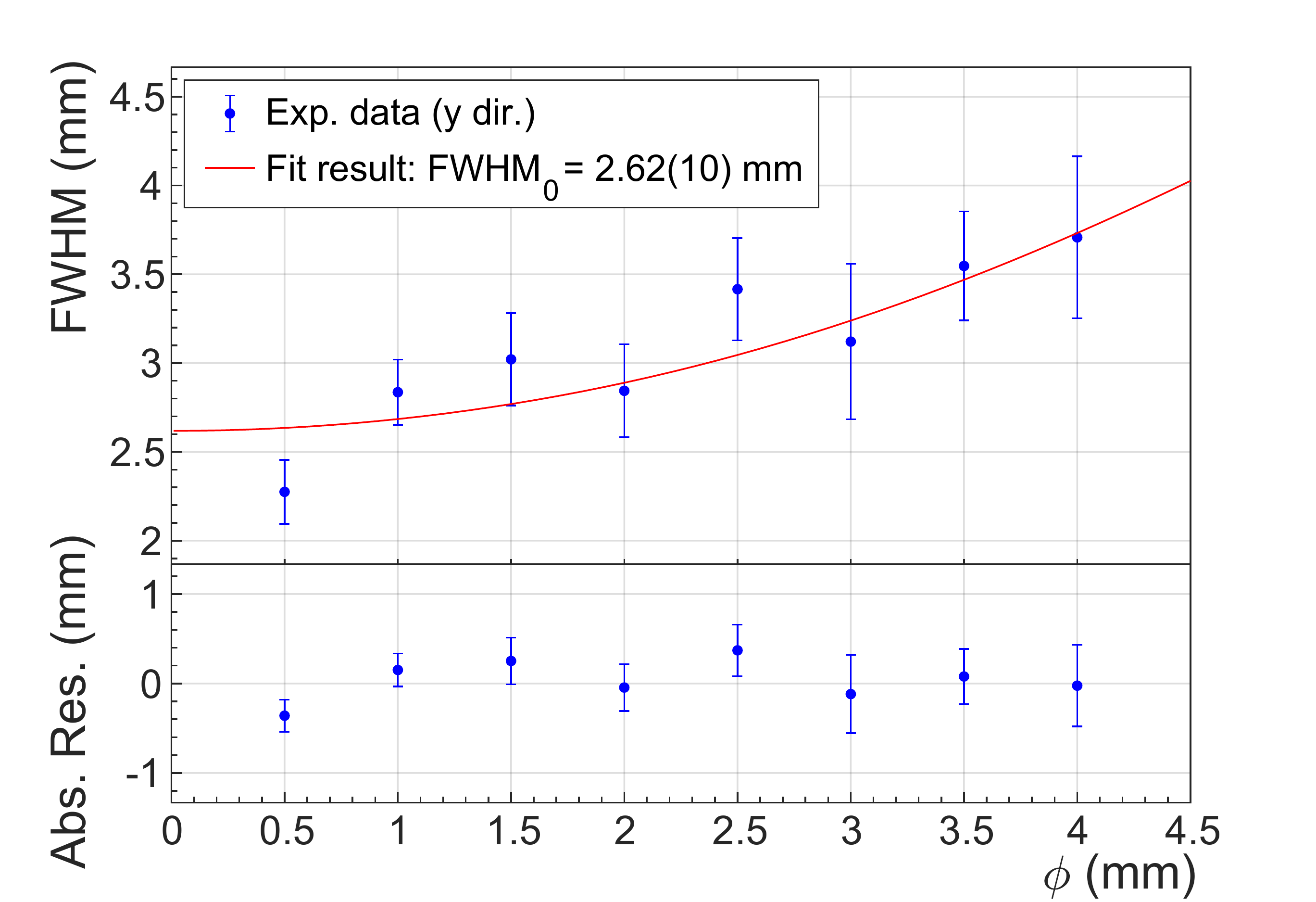}
\end{subfigure}
\caption{Convolution fit for holes with different diameters, resulting in the spatial resolution in the X (left) and Y (right) directions. We also present the absolute residuals (Abs. Res.), i.e. the difference between the experimental data and the fit, for each plot. Figures adapted from \cite{LucasMaster}.}
\label{fig:convFits}
\end{figure}

The second method relies on the fact that the FWHM of an ideally punctual hole converges to the spatial resolution of the detector, which is the width of the point spread function (PSF). The projection $P_{meas}$ obtained for a hole with a given diameter is the convolution of the PSF, for which it is safe to assume a 2D Gaussian function, with the projection of a circular area. This projection over the X direction can be written as
\begin{equation}
    P_{meas}(x) = (P_x \circledast g)(x):=\int_{-\infty}^{\infty} f(\tau) g(x-\tau,\sigma)d\tau,
    \label{eq:convolution}
\end{equation}
where $g$ is the Gaussian function and $P_x$ the projection of a circular area over the X direction. Fitting this convolution for holes of several diameters allows us to estimate the spatial resolution. The convolution shown in Eq.~\ref{eq:convolution} is computed numerically and the fit\footnote{the fitted parameter is the resolution itself, since it is linearly dependent on $\sigma$.} is done using the Least Square Method. The result is shown in Fig. \ref{fig:convFits}.

Although compatible with the \SI{3}{\milli\metre} obtained with the ESF, the values obtained with this method are slightly better. This can be explained by the fact that the FWHM assigned to each diameter is the average of the FWHM for several runs using holes of that diameter.

The values obtained with the two different techniques result in a spatial resolution compatible with at least \SI{3}{\milli\metre}. This spatial resolution is compatible with the obtained by similar detectors \cite{ZHOU2020,ZHOU2021} and is mostly limited by the fact that our detector uses only five electronic channels, which inherently only allows the estimation of the centroid of the track's charge distribution, which is at a certain distance from the neutron capture position. This distance is minimized when the maximum length of the track is truncated, such as in the case of  \SI{2}{\milli\metre} drift. It would be possible to strongly enhance the spatial resolution by operating the detector in the  $\mu$TPC mode (such as in \cite{Pfeiffer_2016}), however, it would be necessary to use a much more sophisticated acquisition system.

\subsection{Detector efficiency and gain stability}
\subsubsection{Analytic Calculation}
It is possible to obtain the expected detection efficiency $\varepsilon_d$ from analytical calculations. The first step is to compute the capture efficiency $\varepsilon_c$, which depends on the macroscopic cross-section $\Sigma$ for the converter material and its thickness $d =$ \SI{2.2(1)}{\micro\metre}, where the uncertainty should comprise eventual irregularities of the deposition process. Taking the converter $(^{10}\mathrm{B}_4\mathrm{C})$ density $\rho=$  \SI{2.3}{\gram\per\centi\meter\cubed} \cite{Hoglund_depo}, its neutron capture cross section for \SI{41.8}{\milli\electronvolt} neutrons \mbox{$\sigma =$ \SI{2994(9)}{b}} \cite{tendl}, its molar mass $\mathrm{M}_{\mathrm{B}_4\mathrm{C}} = $ \SI{52}{\gram\per\mol}, and 
the stoichiometry $a$ of $^{10}$B in a \Bcarb converter layer, which is \num{4}, one can obtain the number of target nuclei of the converter: 
\begin{equation}
    N = a\frac{\rho}{\text{M}}N_A = 1.065 \times 10^{23} \, \text{cm}^{-3}.
    \label{eq:N_target_nuclei}
\end{equation}

The result of Eq.~\ref{eq:N_target_nuclei} is used to compute the   macroscopic cross-section \cite{HUSSEIN20071} for the converter layer:
\begin{equation}
    \Sigma = N\sigma = (3.19 \pm 0.01)\times10^{-2}\, \mu\text{m}^{-1}.
\end{equation}

The neutron beam intensity after crossing our converter layer is
\begin{equation}
    I_{\text{beam}} = e^{-\Sigma d} = (93.2 \pm 0.3) \%.
    \label{eq:beamintensity}
\end{equation}
Therefore, the capture efficiency is given by the percentage of absorbed neutrons: 
\begin{equation}
    \varepsilon_c = 1 - e^{-\Sigma d} = (6.8\pm 0.3)\%.
    \label{eq:conversionefficiency}
\end{equation}

\begin{figure}[]
    \centering
    \includegraphics[width=0.8\textwidth]{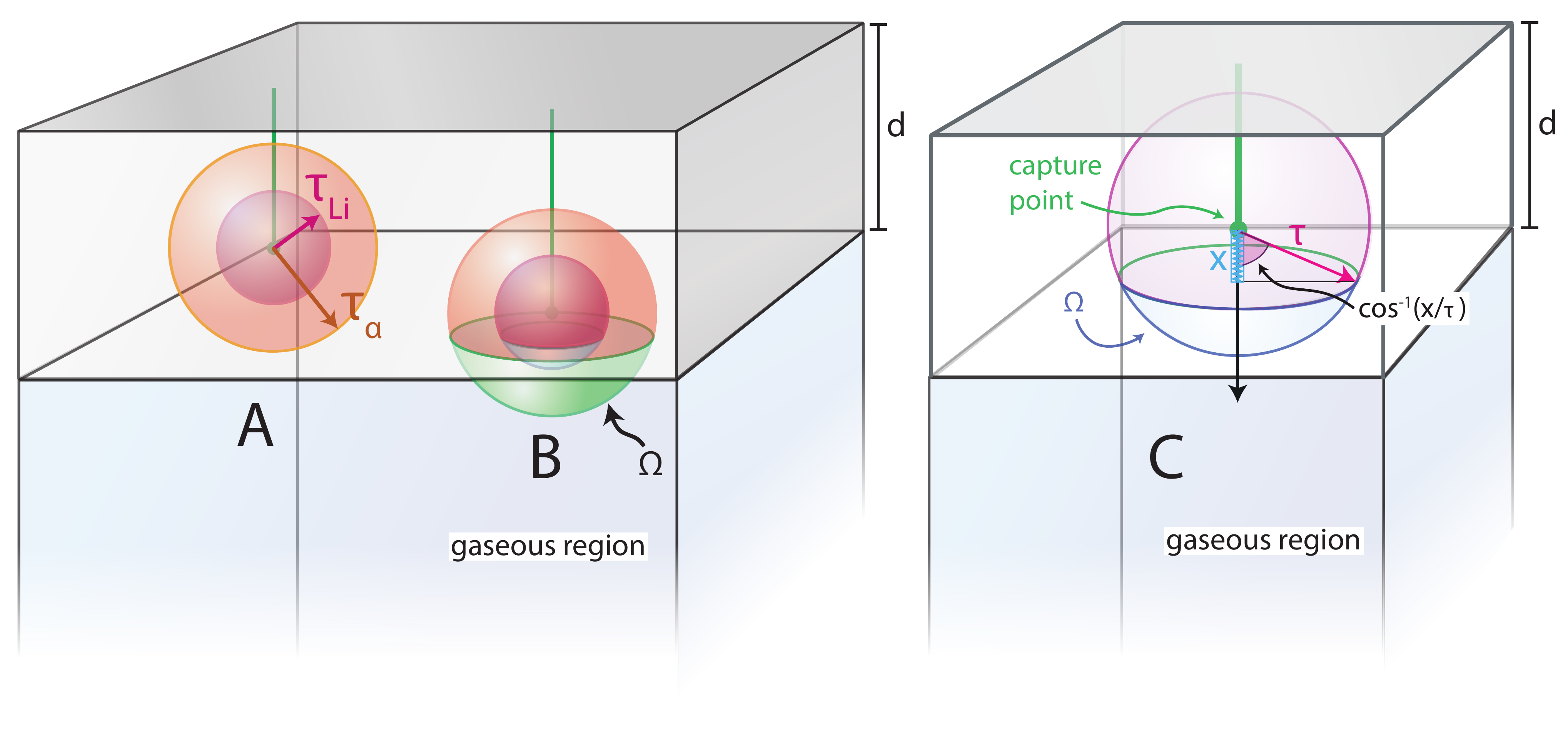}
    \caption{The spheres of radius $\tau_{\alpha}$ and $\tau_{Li}$ define the accessible region for the $\alpha$ and \LiSete nuclei, respectively, inside the converter layer (A). Given certain capture point, only the products within the solid angle $\Omega$ reaches the gas (B). The solid angle $\Omega$ is a function of the angle $\text{cos}^{-1}(x/\tau)$, where $x$ is the smallest distance between the capture point and the drift region (C). Figure adapted from \cite{LucasMaster}.}
    \label{fig:layer_geometricC}
\end{figure}

When the neutron is captured too far from the lower side of the converter layer (at the boundary with the gaseous region), it might not be detected because the products of the capture reaction do not have enough energy to exit the layer, as pictured in Fig. \ref{fig:layer_geometricC} (case A). Therefore, such products have to fulfill a set of geometrical requisites that depend on the capture's position, which are shown in Fig. \ref{fig:layer_geometricC} (case B and C). Finally, the products that enter the drift region with too low energy are also not detected since there is an energy threshold to handle electrical noise and other acquisition system specifications.

\begin{figure}[]
    \centering
    \includegraphics[width=0.8\textwidth]{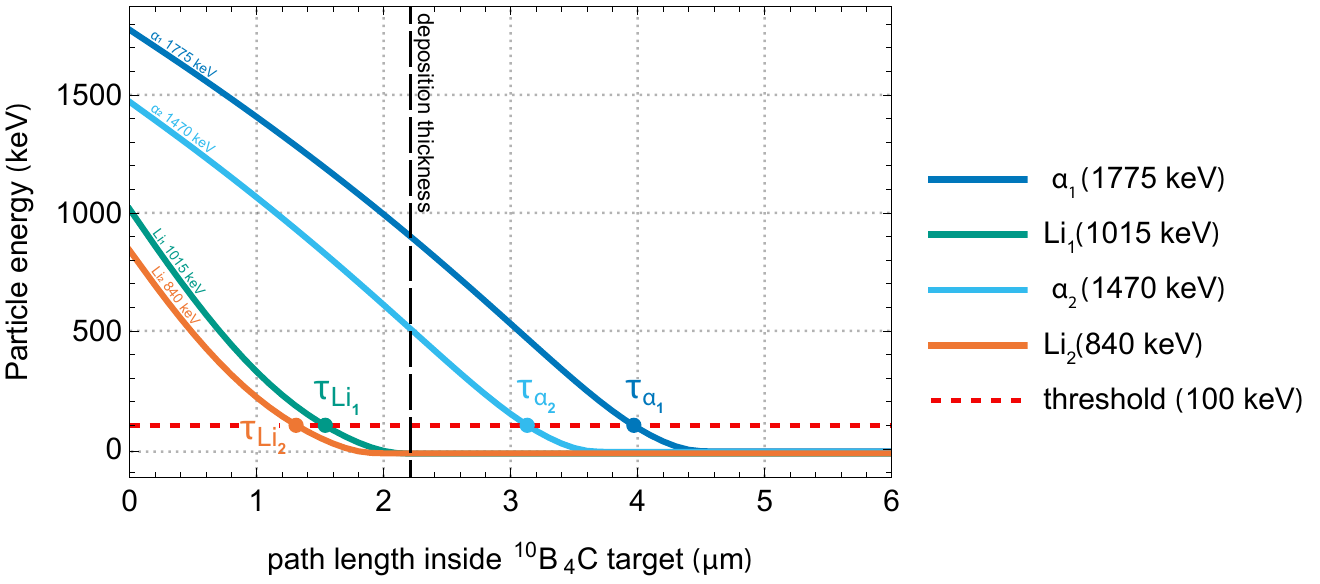}
    \caption{Remaining energy for each \B(n,$\alpha$)\LiSete reaction product as a function of depth for a theoretically infinite \Bcarb converter layer. The $\tau$ values used for the calculations are the depth at which the particle's remaining energy equals the threshold. This condition gives: $\tau_{\alpha_{1}}$(\SI{1775}{\kilo\electronvolt}) $=$ \SI{3.97}{\micro\metre}, $\tau_{\alpha_{2}}$(\SI{1470}{\kilo\electronvolt}) $=$ \SI{3.13}{\micro\metre}, $\tau_{\mathrm{Li_{1}}}$(\SI{1015}{\kilo\electronvolt}) $=$ \SI{1.54}{\micro\metre} and $\tau_{\mathrm{Li_{2}}}$(\SI{840}{\kilo\electronvolt}) $=$ \SI{1.31}{\micro\metre}.}
    \label{fig:Eremainingplot}
\end{figure}

In order to account for the constrains discussed above, the values for $\tau_{\alpha}$ and $\tau_{\text{Li}}$, for both possible decay channels of Eq.~\ref{eq:decays}, were evaluated from direct integration of the average dE/dx plot generated through SRIM \cite{ZIEGLER20101818}.  It is also necessary to consider the energy threshold, which was done by finding $\tau$ for which the average remaining energy for the given particle is equal to the threshold used in the experiment, of \SI{100(10)}{\kilo\electronvolt}. This process is shown in Fig. \ref{fig:Eremainingplot}.

The considerations above are translated to the calculation for the detection efficiency $\varepsilon_d$ as follows:

\begin{equation}
    \varepsilon_d = F\varepsilon_c = F(1 - e^{-\Sigma d}),
    \label{eq:deteff1}
\end{equation}
where $F$ is the share of capture products emitted within the solid angle $\Omega$ of Fig. \ref{fig:layer_geometricC}B, and is computed as

\begin{equation}
    F = \sum_{\text{all products}}F(d,\tau_i) = \sum_{\tau_i=\tau_{\alpha_{1}},\tau_{\alpha_{2}},\tau_{\mathrm{Li}_{1}},\tau_{\mathrm{Li}_{2}}}\int_{0}^{d}\int_{0}^{\text{cos}^{-1}(x/\tau_i)}d\Omega dx \approx 0.477,
    \label{eq:detectablefraction}
\end{equation}
for a boron carbide layer of thickness \SI{2.2}{\micro\metre}, where $\tau_{i}$ are the path length for each product: $\tau_{\alpha_{1}} = $ \SI{3.97}{\micro\metre}, $\tau_{\alpha_{2}} = $ \SI{3.13}{\micro\metre}, $\tau_{\mathrm{Li_{1}}} = $ \SI{1.54}{\micro\metre} and $\tau_{\mathrm{Li_{2}}} = $ \SI{1.31}{\micro\metre}. As result, we obtain the neutron detection efficiency 
\begin{equation}
    \varepsilon_d = (3.22 \pm 0.14)\,\%.
    \label{eq:deteff_final}
\end{equation}

\subsubsection{Simulations}
Monte Carlo simulations play an important role in the development and optimization of this kind of detector. For this purpose, GEANT4 is a powerful tool to describe the detector and for the simulation of the interaction of particles with matter \cite{Agostinelli2003}, allowing the simulation of the nuclear reaction, like Eq.~\ref{eq:decays}, as well as the transport of the reaction products inside the detector.

The detector was simulated with GEANT4 using the physics list \textit{QGSP\_BERT\_HP} (which has high precision models for low energy neutrons \cite{geant4_physicsReference}), according to the device described in section \ref{sec:experimental}. The simulation consisted in a parallel uniform beam with \num{e6} neutrons of \SI{1.399}{\angstrom} ($41.80\, \mathrm{meV}$), hitting the detector in a direction perpendicular to the converter layer. 

\begin{figure}[H]
    \centering
    \includegraphics[width=0.575\textwidth]{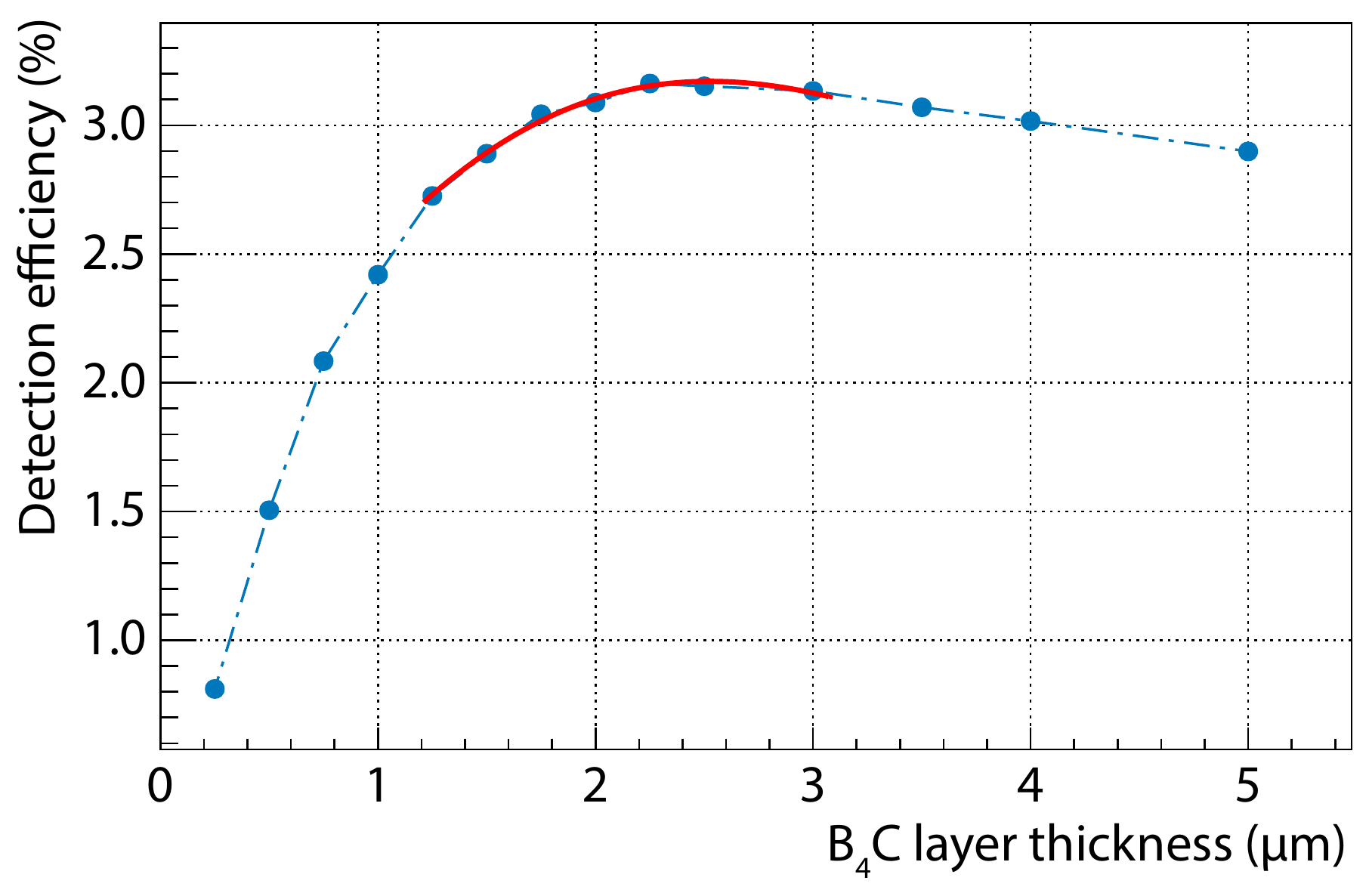}
    \caption{Detection efficiency for one layer of boron carbide, simulated with GEANT4. The fit for maximum efficiency is an arbitrary function adjusted just to guide the eye within the interval that comprises the data around the peak (continuous line).}
    \label{fig:sim_eff}
\end{figure}

Taking 14 different thicknesses ranging from $\SI{0.25}{\micro\meter}$ to $\SI{5}{\micro\meter}$ of the boron carbide converter layer, we evaluate the thermal neutrons detection efficiency by counting the charged particles that leave the boron converter and reach the gaseous volume. For a nominal thickness of $\SI{2.2}{\micro\meter}$, an efficiency of $\SI{3.14(10)}{\percent}$ was obtained. While the maximum efficiency was achieved with a thickness of $\SI{2.5}{\micro\meter}$ resulting in $\SI{3.17(10)} {\percent}$, as shown in Fig. \ref{fig:sim_eff}. We did not considered the experimental threshold (\SI{100}{\kilo\electronvolt}) when evaluating the theoretical efficiency in this case.

\begin{figure}[H]
    \centering
    \includegraphics[width=0.6\textwidth]{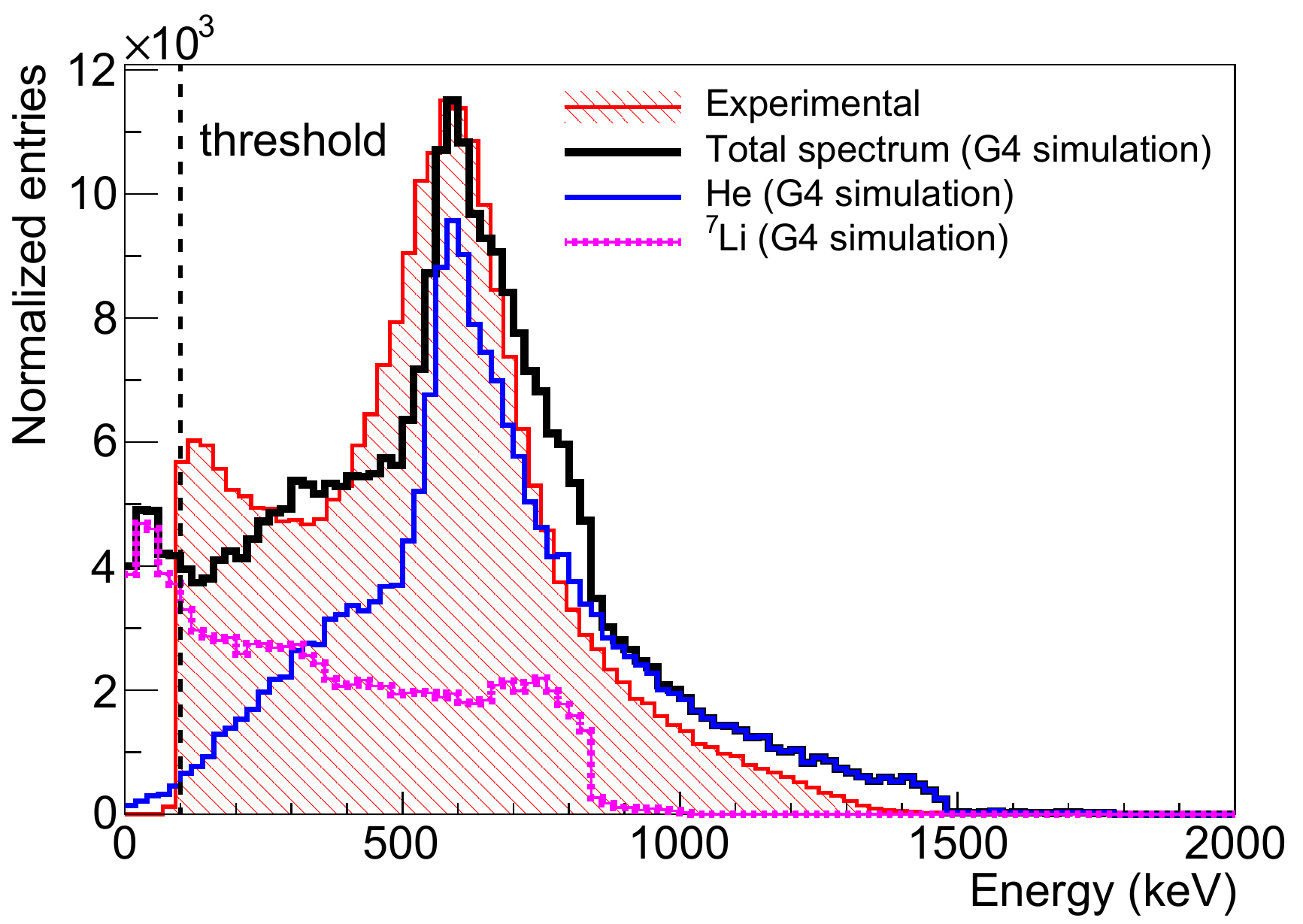}
    \caption{Experimental energy spectrum compared with the GEANT4 simulation. The simulation does not include the effect of the detector finite energy resolution and was normalized to the maximum value of the experimental energy distribution. Figure from \cite{LucasMaster}.}
    \label{fig:sim_spectra}
\end{figure}

Using the low-energy electromagnetic model Livermore \cite{geant4_physicsReference}, we simulated the deposited energy by the reaction products inside the $\SI{2}{\milli\meter}$ of the drift region. This simulation is used as an energy pre-calibration for the experimental spectrum since both present the same primary structure, consisting of an energy peak around \SI{600}{\kilo\electronvolt}, as shown in Fig. \ref{fig:sim_spectra}. We did not include the experimental energy threshold (\SI{100}{\kilo\electronvolt}) of our system nor the charge multiplication process in the simulation, which means considering that the acquired energy is equal to the energy deposited by the reaction products in the absorption region of the detector. 

\subsubsection{Experimental measurements}
The detection efficiency is evaluated by dividing the number of detected neutrons $N_{det}$ by the total number of neutrons $N_{tot}$ hitting the detector. We calculated the total number of neutrons from the neutron flux $\phi$, measured as described in section \ref{sec:experimental}, the area $A$ selected from the beam and the time length $\Delta t$ of the measurement. The neutron flux was corrected using the counts of a fission chamber installed at the neutron beam outlet using as reference the number of counts registered by this same chamber during the neutron flux calibration process. With these values, the experimental neutron detection efficiency is given by

\begin{equation}
\label{eq:deteff}
    \varepsilon = \frac{N_{det}}{N_{tot}} = \frac{N_{det}}{{\phi A \Delta t}}.
\end{equation}

Using cadmium masks with squared holes, we selected different areas from the beam: \SI{0.36}{\centi\metre\squared} and \SI{1.44}{\centi\metre\squared}, at the same position where the flux measurement was taken. The total number of detected neutrons $N_{det}$ was counted with negligible uncertainties at these rates. We obtained \SI{2.75(26)}{\percent} detection efficiency for the \SI{0.36}{\centi\metre\squared} mask and \SI{2.57(15)}{\percent} for the \SI{1.44}{\centi\metre\squared}, which are compatible within $1\,\sigma$ with each other.

The gain stability was monitored in real-time, in order to provide necessary corrections to the image reconstruction in offline data analysis. A gain shift would imply changes in efficiency since the energy spectrum would be displaced from its original position with respect to the energy threshold. Besides that, changes in the gain over time may bring the detector to an unstable operation in terms of discharge probability.

\begin{figure}[H]
  \centering
  \includegraphics[width=0.6\linewidth]{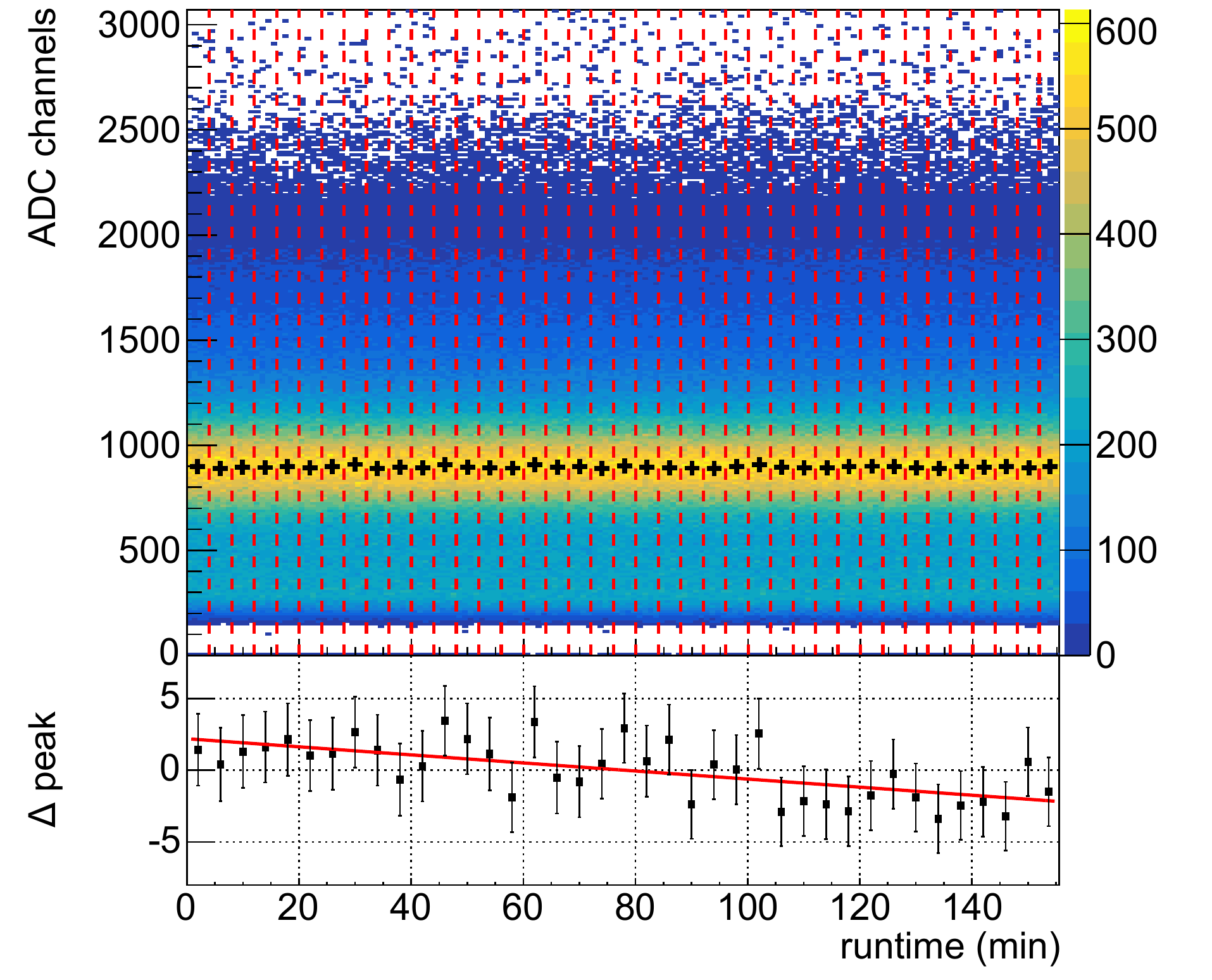}
  \caption{Temporal stability analysis with the tendency analysis. This run has 4793145 events and was divided into batches of \SI{4}{\minute} (\num{123403} entries each), generating 39 peaks. The average of the peak position was \num{897(2)} ADC channels. The fit for $\Delta_{\text{peak}}$ resulted an angular coefficient $-0.028 \pm \, 0.009$ ADC channel/min and a linear coefficient equals to  $2.2 \pm\, 0.8$ ADC channels.}
  \label{fig:TimeEvo}
\end{figure}

The data shown in Fig. \ref{fig:TimeEvo} present a method to analyze the gain stability: the energy channel of the run was divided into batches of \SI{4}{\minute} each (separated by the dashed vertical lines) and plotted over time. Then, we locate the spectrum peak using a peak identification algorithm \cite{MORHAC2000108} and fit a Gaussian to it, registering its center as the cross marker in Fig. \ref{fig:TimeEvo}. Slow gain changes due to environmental changes are easily seen as a tendency in the sequence of the points, whose energy spectrum can be drifted by a reasonable amount as it happens in the case reported by \cite{geovoXRF}. The tendency analysis can be carried out by fitting the behaviour of the difference between each peak position and the average position of all peaks, which we labelled as $\Delta_{\text{peak}}$ in the Fig. \ref{fig:TimeEvo}.

There is a slight drift of the spectrum's peak within about \SI{2.5}{\hour}, as the fit result of Fig. \ref{fig:TimeEvo} suggests. However, this drift is smaller than \num{10} ADC channels, corresponding to about \SI{0.2}{\percent\per\hour} change, which is not statistically sufficient to be interpreted as a reasonable change. Also, at the beginning of the operation, some slow gain stabilization of these detectors are expected, as reported in other works \cite{CHATTERJEE2021165749, Chatterjee_2020, HAUER2020164205, ALFONSI20126, CHERNYSHOVA2020111755}. The monitoring remains essential since it can be used to correct the detector's efficiency if significant changes happen, using the detector's current rate and the neutron monitor fission chamber.

\section{Final Considerations}
A neutron detector was built with a single boron carbide layer and a resistive charge division readout. This detector shows the feasibility of a position-sensitive detector that can be easily mounted in a neutron beam line based on simple standard electronics. The detector also operates above of the background gamma rays sensitive region (registering no events when the neutron beam is off), an essential feature for its use in environments like nuclear reactor research facilities. If needed, the gamma sensitivity can be controlled by changing the drift region, and a new energy pre-calibration can be quickly obtained with new simulation results. 

The average of the obtained thermal neutron detection efficiency is \SI{2.66(15)}{\percent}, which agrees within 2.7 $\sigma$ with the prediction of \SI{3.22(14)}{\percent} and with the preliminary simulation of \SI{3.14(10)}{\percent}, and can be reduced by altering the layer thickness allowing different applications, such as beam monitors.
At the current configuration, one can use the detector to measure the beam profile given that the position resolution obtained was at least \SI{3}{mm}. There is also the possibility to raise the efficiency by adding new converter layers over the GEMs, as seen in other works \cite{MBGEM,KLEIN2011, KOHLI2016}. For future works and following that same idea, it is possible to use boron coated Thick-GEMs, operating with unitary effective gain, raising the thermal neutron detection efficiency.

\acknowledgments
This work was supported by CNPq grant 156767/2019-8.\\
H. Natal da Luz acknowledges project GAČR GA21-21801S (Czech Science Foundation).

\bibliography{./lib/bibliography.bib}
\bibliographystyle{./lib/JHEP}

\end{document}